\documentclass[aps, twocolumn, pra, showkeys, 10pt]{revtex4}
\usepackage{amsmath}
\usepackage{graphicx}
\begin{document}
\title{High-stability time-domain balanced homodyne detector for ultrafast optical pulse applications}
\author{Merlin Cooper}
\email{m.cooper1@physics.ox.ac.uk}
\author{Christoph S\"{o}ller}
\author{Brian J. Smith}
\affiliation{Clarendon Laboratory, University of Oxford, Parks Road, Oxford, OX1 3PU, UK}
\begin{abstract}
Low-noise, efficient, phase-sensitive time-domain optical detection is essential for foundational tests of quantum physics based on optical quantum states and the realization of numerous applications ranging from quantum key distribution to coherent classical telecommunications. Stability, bandwidth, efficiency, and signal-to-noise ratio are crucial performance parameters for effective detector operation. Here we present a high-bandwidth, low-noise, ultra-stable time-domain coherent measurement scheme based on balanced homodyne detection ideally suited to characterization of quantum and classical light fields in well-defined ultrashort optical pulse modes.
\end{abstract}
\maketitle
\section{Introduction}

\noindent Complete characterization of the amplitude and phase of a single optical field mode is a key requirement for a range of applications, both quantum and classical, such as quantum state discrimination \cite{Muller:12, Wittmann:08}, quantum state and process tomography \cite{Smithey:93, Lvovsky:09, Lobino:08}, quantum communications \cite{Grosshans:03}, quantum-enhanced reading of a classical memory \cite{Pirandola:2011}, and ultrasensitive linear sampling of fiber optical systems \cite{Dorrer:03}. Balanced homodyne detectors (BHDs) offer a straightforward approach to measure both amplitude and phase information of a single optical field mode. BHDs measure the quadratures of a well-defined electromagnetic field mode with high efficiency and minimal technical noise \cite{Yuen:83}. Significant effort over the past decade has focused on increasing the bandwidth, efficiency and signal-to-noise characteristics of BHDs to enable coherent detection of pulsed optical modes \cite{Hansen:01, Zavatta:02, Zavatta:06, Okubo:08, Haderka:09, Huisman:09, Chi:11,Kumar20125259}. This enables not only significantly increased data acquisition rates, avoiding experimental drift, but more fundamentally it allows time-domain characterization of conditionally-prepared quantum states and conditional processes \cite{Lvovsky:09, Lobino:08, Zavatta:06}. A key challenge that has only recently come to light \cite{Zavatta:02, Haderka:09} that we address here is the detector stability, which is necessary to measure quadrature distributions that are not symmetric about the origin.

Bandwidth, noise, efficiency, and stability are the key performance properties of a BHD. The detector capability to distinguish individual optical pulses, corresponding to the detector bandwidth, directly impacts the accuracy of the measured quadratures. Non-classical quantum states of light can exhibit quadrature distributions with fine, non-Gaussian structure \cite{leonhardt}. Excess noise and non-unit efficiency lead to smoothing of the measured quadrature distributions \cite{Appel:07}, which can lead to the inability of the BHD to resolve such quantum signatures. Furthermore, low-frequency temporal instability results in a shift of the detector baseline, corresponding to a drift in the quadrature origin. This instability significantly impacts the ability to measure optical states that are not symmetric about the origin of the quadrature phase space as well as the capacity to discriminate closely separated states in phase shift keying protocols \cite{Chi:11,Muller:12}. Detector instability is often circumvented by continuously calibrating the detector, or by making assumptions about the nature of the state of the field, e.g. that it is symmetric about the origin. Although this approach may be used in special circumstances, measurement of an arbitrary unknown state of light requires a detector with good response to the input state as well as stability. These detector requirements can be combined into two helpful parameters to describe the overall detector performance, namely the signal-to-noise ratio (SNR) and the time-bandwidth product (TBP), the latter we introduce for the first time to characterize a BHD.

In this article we report a BHD with an unprecedented combination of SNR and TBP, with an electronic bandwidth of $80$ MHz, able to perform shot-noise-limited measurements of ultrafast optical field quadratures. The detector has a $14.5$ dB signal-to-noise ratio, which surpasses the best achieved to date at this bandwidth \cite{Kumar20125259}. Time-resolved field quadrature measurement of an $80$ MHz optical pulse train is confirmed by calculating the correlation coefficient between adjacent temporal modes. Measurement of the detector Allan deviation \cite{Allan:66} indicates the detector is stable for measurement times up to approximately 2 seconds, which is over five orders of magnitude greater than any reported \cite{Zavatta:02, Haderka:09}. This stability enables measurement of $160$ million quadrature values between calibrations at the laser repetition rate. This gives a high time-bandwidth product of $\Delta f \Delta t = 1.6 \times 10^8 $, where $\Delta f$ is the detector bandwidth and $\Delta t$ is the stability interval. Finally, to demonstrate the detector performance in the quantum domain, state reconstructions of a weak coherent state and heralded single-photon state occupying a femto-second pulsed optical mode are presented.

\section{Detector design}

\begin{figure}
\includegraphics[width=1.0\linewidth]{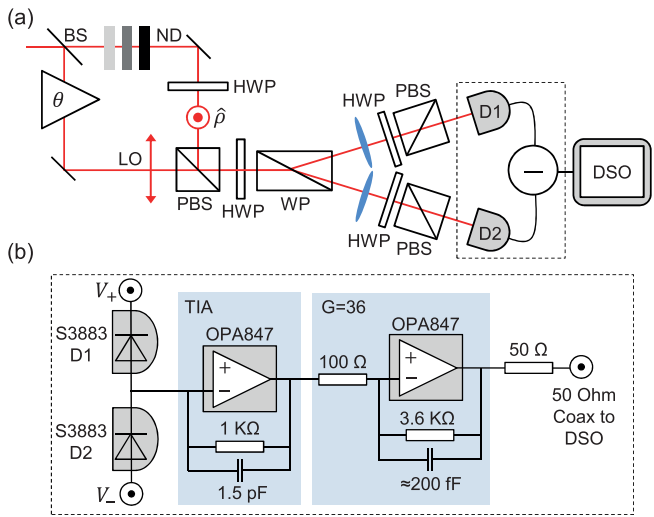}
\caption{(a) Electro-optic schematic of BHD as described in text. (b) BHD electronic circuit schematic showing the main components: photodiodes D1 and D2 wired in series, transimpedance amplifier (TIA) and additional voltage gain stage (G=36).}
\label{fig:setup}
\end{figure}

In balanced homodyne detection a signal field mode with unknown state $\hat\rho$, to be characterized, is interfered with a matched reference mode called the local oscillator (LO) on a 50:50 beam splitter, Fig.~\ref{fig:setup} (a). The two output modes are detected with square-law detectors. The difference photocurrent yields a direct measure of the generalized quadrature $\hat{X_{\theta}}=\hat{X}\cos\theta+\hat{P}\sin\theta$ of the signal field in state $\hat{\rho}$, where $\theta$ is the relative optical phase between the signal and LO \cite{Lvovsky:09}. Repeated measurements of $\hat{X_{\theta}}$ for an ensemble of identically prepared states enables estimation of the conditional quadrature probability density $\text{Pr}(X_{\theta}|\hat{\rho})=\langle X_{\theta} | \hat {\rho} | X_{\theta} \rangle$, where $|X_{\theta}\rangle$ is the quadrature eigenstate with eigenvalue $X_{\theta}$ \cite{Lvovsky:09}. 

A schematic representation of the detector is shown in Fig.~\ref{fig:setup} (a). The LO and signal beams consist of $100$ fs optical pulse trains ($830$ nm central wavelength) derived from a mode-locked Ti:Sapphire laser oscillator ($80$ MHz repetition rate). The relative phase is monitored using a diode laser at $1550$ nm (not shown). The signal can be prepared in either a coherent state with amplitude controlled using calibrated neutral density (ND) filters, or a heralded single-photon state generated by spontaneous parametric down conversion \cite{Mosley:08}. The signal and LO are combined into a single spatial mode on a polarizing beam splitter (PBS). A half-wave plate (HWP) and Wollaston prism (WP) act as a precise 50:50 beam splitter (BS), with output modes focused onto two p-i-n photodiodes (D$1$($2$), Hamamatsu S3883) operating in photo-conductive mode. In front of each diode a HWP and PBS provide variable loss to compensate for different quantum efficiencies. The photodiodes are wired in series to produce the difference current, which is converted to a voltage by a transimpedance amplifier (TIA) and subsequently amplified with gain $G=36$, as depicted in Fig.~\ref{fig:setup} (b). This gives a combined transimpedance gain of $36$ kV/A. The amplifier stages utilize Texas Instruments OPA$847$ operational amplifiers. The output voltage signal is digitized by a Tektronix MSO$5104$ digital storage oscilloscope (DSO) which we computer control from MATLAB over a TCP/IP interface.

Careful attention must be paid both to the design and construction of the electronic circuit and the optical setup in which the BHD is employed. The layout of the printed circuit board (PCB) will influence the performance of the circuit through parasitic capacitance and inductance. To this end, signal tracks are kept as short as possible, and generally at right angles to tracks carrying power supplies to the amplifiers and photodiodes. Top and bottom ground planes are employed to further reduce coupling between adjacent tracks. Surface-mount components are used throughout. The circuit was modeled in SPICE to optimize the values of feedback capacitors and resistors for the two amplifier stages, shown in Fig.~\ref{fig:setup} (b). We found that the response predicted by the model very closely matched the measured response of the detector in terms of electronic pulse duration, noise performance and overshoot. The model is used to select the optimal component values for a target detector bandwidth and gain.

The detector response will be affected by the spot size of the focused local oscillator mode on the photodiode active area. A tightly focused mode will cause the photodiode to saturate at high pulse energies, thus making the photodiode response non-linear. Care was taken to select two photodiodes exhibiting similar temporal response characteristics and quantum efficiencies. Inhomogeneities in the photodiode quantum efficiency as a function of beam position will lead to instability of the detector balance if the laser pointing stability is poor, as will poorly regulated photodiode bias supplies. Thus the stability of the detector is governed by a combination of environmental and electronic effects.

\section{Detector characterization}
\begin{figure}
\includegraphics[width=1.0\linewidth]{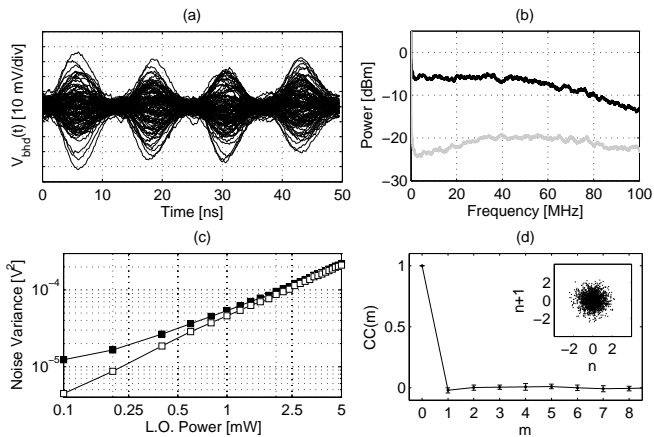}
\caption{(a) Time-domain detector output voltage $V_{\text{bhd}}(t)$ showing pulsed shot noise of vacuum. (b) Frequency spectrum of detector electronic noise (grey) and shot noise at 5 mW L.O. power (black). (c) Detector noise variance versus L.O. power: black (white) squares: with electronic noise (subtracted). (d) CC between different pulses with parametric plot of 2000 vacuum quadrature samples for pulse $n$ and $n+1$ (inset).}
\label{fig:noise_variance_corr}
\end{figure}

In this section we introduce and present measurement results for a number of different parameters to assess the detector bandwidth, signal-to-noise ratio and stability. Time-traces of the BHD output voltage $V_{\text{bhd}}(t)$ for vacuum state input are shown in Fig.~\ref{fig:noise_variance_corr} (a). The full-width half-maximum of a single electrical pulse is $5.5$ ns, well below the laser inter-pulse time separation. One measured quadrature sample $X_{\theta}$ of the signal in state $\hat{\rho}$ is related to the detector output voltage $ V_{\text{bhd}}(t)$ through
\begin{equation}
X_\theta=\frac{\int_t^{t+\tau_p}V_{\text{bhd}}(t')dt' - \int_t^{t+\tau_p} V_{\text{residual}}(t')dt'}{\sqrt{2}\eta G e|\alpha_{\text{LO}}|}, 
\label{eq:bhd_voltage}
\end{equation}
where $\tau_p$ is the duration of one electronic pulse, $\eta$ is the quantum efficiency of the photodiodes, $e$ is the elementary charge, $G$ is the net transimpedance gain of the amplifier stages, $|\alpha_{\text{LO}}|$ is the local oscillator amplitude and $V_{\text{residual}}(t)$ is the residual detector output voltage due to imperfect photocurrent subtraction and D.C. offset. The expression in Eq.~(\ref{eq:bhd_voltage}) is evaluated over a set of signal pulses in state $\hat{\rho}$ to obtain a set of quadrature samples $\left\{X_{\theta}\right\}$, from which $\text{Pr}(X_{\theta}|\hat{\rho})$ is estimated. The scale factor $\sqrt{2}\eta e G |\alpha_{\text{LO}}|$ and baseline offset $\int_t^{t+\tau_p} V_{\text{residual}}(t')dt'$ are uniquely determined by acquiring quadrature samples of the vacuum state $\{ X_{\theta}\}$ \cite{Appel:07} which satisfy $\langle {X}_{\theta}\rangle=0$ and $\text{Var}\left( {X}_{\theta}\right)=\frac{1}{2}$. This process constitutes calibration of the detector.

The voltage noise variance $\text{Var}\left(  \int_t^{t+\tau_p}V_{\text{bhd}}(t')dt' \right)$ of a BHD for vacuum state input is expected to scale linearly in the LO power with a constant offset representing the detector electronic noise, and second-order deviation due to fluctuations in the detector balance \cite{Chi:11}. The measured detector noise variance is shown in Fig.~\ref{fig:noise_variance_corr} (c) for a range of LO powers. Each point is calculated from $40,000$ pulses with the signal mode blocked. We observe a linear dependence confirming the detector is shot-noise limited over a wide range of LO powers. The signal-to-noise ratio is $14.5$ dB at 5 mW LO power, which represents an effective quantum efficiency due to electronic noise $\eta_{\text{en}}=0.96$ \cite{Appel:07}. When combined with the photodiode efficiency $\eta_{\text{pd}}=0.90$ this gives an overall detector efficiency $\eta_{\text{bhd}}=\eta_{\text{en}}\eta_{\text{pd}}=0.86$, which is well above the threshold necessary to observe quantum features \cite{Lvovsky:09}. The photodiode efficiency could be increased to approximately $\eta_{\text{pd}}=0.97$ by removing the input window. 

The detector bandwidth is determined from the noise spectrum, depicted in Fig.~\ref{fig:noise_variance_corr} (b). The black line is the detector response at $5$ mW LO power with vacuum input, while the grey line is the detector electronic noise with no LO. Note the -3 dB point in the shot noise at approximately 80 MHz indicating the electronic bandwidth of the detector. As a further measure that the detector can temporally resolve pulses at the laser repetition rate, we calculate the correlation coefficient (CC) between subsequent vacuum quadrature measurements defined as
\begin{equation}
CC(m)=\frac{\langle X_n X_{n+m} \rangle - \langle X_n \rangle \langle X_{n+m} \rangle}{{\sqrt{\text{Var}({X_n})}}{\sqrt{\text{Var}({X_{n+m})}}}},
\end{equation}
where $\langle ... \rangle$ indicates an average over the measurement ensemble, $X_n$ and $X_{n+m}$ are quadrature samples of pulses separated by $m$ periods of the master laser and $\text{Var}({X_n})$ is the variance of the quadrature statistics in the ensemble required for normalization of the CC. Fig.~\ref{fig:noise_variance_corr} (d) shows the CC as a function of $m$ for the normal working LO power of 5 mW for $m=0...9$. The uncertainties in each value are given by the standard deviation of the CC from 20 separate measurements and indicate the random statistical fluctuations in $CC(m)$. The CC between pulse $n$ and  $n+1$, $CC(1)=-0.019\pm 0.02$, which is comparable with other reported BHDs \cite{Chi:11,Okubo:08}, and suggests that a repetition rate even higher than 80 MHz could be used. There is no significant correlation between the measured quadrature value of pulse $n+1$ from that of pulse $n$, also represented by the lack of correlation in the parametric plot, inset Fig.~\ref{fig:noise_variance_corr} (d). Confirmation of the detector bandwidth and evaluation of the CC thus demonstrates that our detector is capable of shot-noise-limited measurement of conditional states and processes at an 80 MHz repetition rate. 

The common mode rejection ratio (CMRR) represents the detector capacity to efficiently subtract amplitude fluctuations of the two beams incident on the photodiodes. This can be determined by measuring the output voltage for a vacuum input with the detector balanced and with one photodiode blocked. At the laser repetition rate the CMRR is $63$ dB with $5$ mW LO power. The CMRR is maximized by independently adjusting the photodiode bias voltages ($\text{V}_{+(-)}$ in Fig.~\ref{fig:setup} (b)) using well-regulated fine-adjust DC power supplies and tuning the relative arrival time of the two optical modes on each photodiode. To assess the photodiode linearity in response to the $100$ fs pulses, the photocurrent produced by one diode before amplification was measured for varying incident optical power. The diodes were found to be linear for average optical intensities of up to $10$ mW per diode. This confirms the detector linearity in the normal working regime of $2.5$ mW per diode.

\begin{figure}[t]
\includegraphics[width=1.0\linewidth]{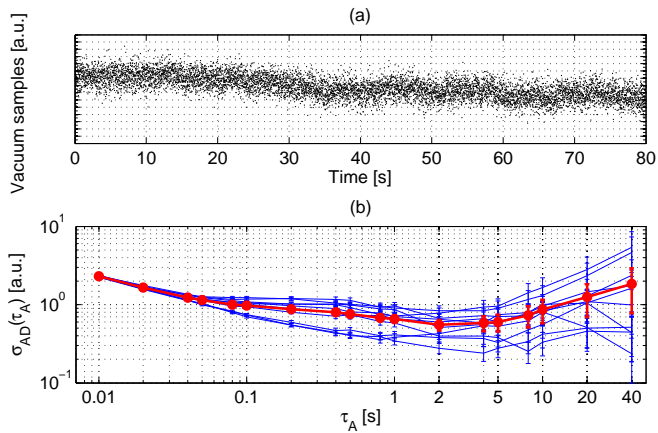}
\caption{Stability characterization: (a) Typical evolution of vacuum state samples as a function of time - a linear drift in the detector baseline is evident over the 80 second sampling window. (b) Allan deviation, $\sigma_{\text{AD}}(\tau_A)$ of measured vacuum samples for ten independent measurements of duration 80 seconds (blue curves) and the average Allan deviation and associated statistical errors (red curve).}
\label{fig:allan}
\end{figure}

In standard operation a BHD is initially calibrated by acquiring quadrature measurements of the vacuum before sampling the quantum state of interest as described above. This calibration enables scaling of the raw detector output voltage  to units of dimensionless quadrature, and sets the detector baseline. However, the calibration is not valid indefinitely due to changes in either the scale factor $\sqrt{2}\eta e G |\alpha_{\text{LO}}|$ or the detector baseline $\int_t^{t+\tau_p} V_{\text{residual}}(t')dt'$ in Eq.~(\ref{eq:bhd_voltage}). Ideally, $ V_{\text{residual}}(t+n\tau)= V_{\text{residual}}(t)$ for $n=-\infty...\infty$, where $\tau$ is the period of the laser. In practice the detector baseline changes over time due to a combination of phenomena such as fluctuations in the photodiode bias voltages, varying beam-splitter ratios, scattering of light from dust particles and laser pointing instability \cite{Stefszky:12}. The effect of changes in the detector baseline on the measured quadrature samples is evidenced in Fig.~\ref{fig:allan} (a) which shows vacuum state samples obtained by the  detector over a time interval of 80 seconds with 5 mW LO power. During this time interval various environmental influences, such as those outlined above, have caused the detector baseline to change in an approximately linear manner with time. Without re-calibration this will lead to errors in state discrimination and tomography applications.

To perform a more quantitative analysis of the drift in Fig.~\ref{fig:allan} (a) we calculate the Allan deviation $\sigma_{\text{AD}}(\tau_A)$ of the vacuum state quadratures \cite{Allan:66,Zavatta:02,Haderka:09}. This statistical measure allows the stability of a parameter on different timescales to be determined (applied originally to clocks). The Allan deviation is defined as
\begin{equation}
\sigma_{\text{AD}}(\tau_A)=\sqrt{\frac{1}{2}\left\langle(\overline{X}_{n+1}-\overline{X}_{n})^2\right\rangle},
\end{equation}
where $\overline{X}_{n+1}$ and $\overline{X}_{n}$ are averages of the vacuum quadratures over adjacent intervals of length $\tau_A$ and $\langle ... \rangle$ denotes averaging over the whole dataset, i.e. as many adjacent intervals of length $\tau_A$ that fit within the overall measurement time of 80 seconds (e.g. for $\tau_A=40 s$ there are only two adjacent intervals in the dataset). 
We collected 10 separate datasets containing vacuum state samples, each of duration 80 seconds. The Allan deviations were computed for each dataset, shown as blue curves in Fig.~\ref{fig:allan} (b). We then averaged these separate Allan deviations to get the typical Allan deviation, shown as the red curve in Fig.~\ref{fig:allan} (b). From this result we conclude that the drift is not significant for measurement times up to approximately 2 seconds, indicated by the minimum in the Allan deviation at $\tau_A=2$ s. This result is intuitively correct from examining the trend in vacuum state samples shown in Fig.~\ref{fig:allan} (a). Based on the rate of drift of the detector baseline it is clear that for data collection periods of up to about 2 seconds there will not be any significant detector drift. We argue that it is important to characterize the stability of a particular BHD setup to determine the frequency of re-calibration required to avoid baseline drift errors in measurements. 

\begin{figure*}
\includegraphics[width=0.7\linewidth]{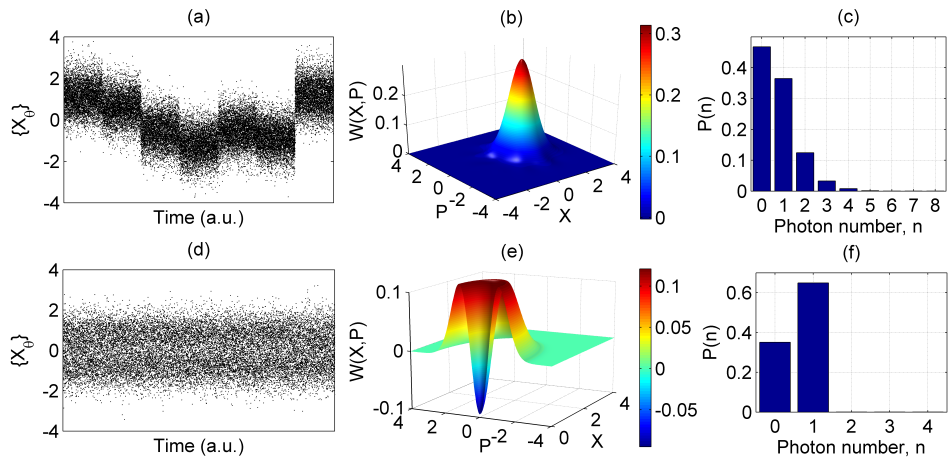}
\caption{Demonstration of quantum state tomography of a coherent state (top row) and single-photon Fock state (bottom row). Left to right: raw quadrature samples $\{X_{\theta}\}$ of the conditional probability densities $\left \{ \text{Pr}(X_{\theta}|\hat{\rho})\right \}_{\theta}$; Wigner functions of the reconstructed states; photon number statistics of reconstructed states.}
\label{fig:wf}
\end{figure*}

Under certain circumstances it is possible to employ notch filters at the laser repetition rate and harmonics to enhance the CMRR; thereby further suppressing the fluctuating detector baseline $\int_t^{t+\tau_p} V_{\text{residual}}(t')dt'$ \cite{Kumar20125259} and potentially improving the stability. The effect of such filtering on the measurement of an unknown state $\hat{\rho}$ is to displace the optical state to the origin of phase space, effectively discarding all information about the displacement amplitude of the state. For measurement of a phase-invariant state, such as a mixture of Fock states, this has no adverse effect. However, for optical states exhibiting coherent amplitude, filtering should be avoided. Alternatively, one can use the fact that the ideal conditional probability densities satisfy the relation $\text{Pr}(X_{\theta}|\hat{\rho})=\text{Pr}(-X_{\theta+\pi}|\hat{\rho})$, where $\text{Pr}(X_{\theta}|\hat{\rho})$ is the inferred conditional probability density at LO phase $\theta$, to elucidate the presence of an offset $\delta X$ in the detector calibration \cite{PhysRevA.85.052129}. However, this technique has limited applicability to the case where the offset $\delta X$ is constant for the duration of acquisition of the whole set of quadrature samples $\{X_{\theta}\}$ from which the probability densities for all LO phases are estimated, $\left \{ \text{Pr}(X_{\theta}|\hat{\rho})\right \}_{\theta}$. Furthermore, one must know the LO phase precisely and independently of the homodyne measurement, thus requiring the use of an auxiliary phase reference. Therefore, in general it is important that the detector be as intrinsically stable as possible when applied in scenarios where there is no prior knowledge of the state being measured.

\section{Quantum state tomography}

One of the primary uses of a shot-noise-limited time-domain balanced homodyne detector is to enable quantum state tomography (QST) of conditionally-prepared quantum states. The acquired quadrature samples $\{X_{\theta}\}$ can be inverted using an iterative maximum-likelihood algorithm \cite{Lvovsky:04} to obtain the density matrix $\hat{\rho}$ of the detected state.

To demonstrate the detector capacity in the quantum domain, we perform QST to reconstruct the Wigner function $W(X,P)$  and photon number statistics $P(n)$ of a weak coherent state ($\hat{\rho}=|\alpha\rangle\langle\alpha|$), Fig.~\ref{fig:wf} (b,c)  and a single-photon Fock state ($\hat{\rho}=|1\rangle\langle 1 |$), Fig.~\ref{fig:wf} (e,f). The coherent state was reconstructed from $35,000$ quadrature samples $\{X_{\theta}\}$ with the relative phase between LO and signal at 7 values in the interval $\theta \in [0,\pi)$, Fig.~\ref{fig:wf} (a). The relative phase between the LO and coherent state $\theta$ was measured using a separate diode laser co-propagating in the interferometer and adjusted using a piezo-mounted mirror. The reconstructed coherent state has a fidelity of 0.99 with a pure coherent state of magnitude $|\alpha|=0.86$ which is consistent with the independently measured detector loss, interference visibility and input laser beam power after attenuation by calibrated neutral density filters. For the heralded single photon, $100,000$ samples $\{X_{\theta}\}$ were measured and the LO phase was allowed to drift freely, Fig.~\ref{fig:wf} (d). The single photon state, which has been corrected for the detector efficiency $\eta_{\text{bhd}}=0.86$, has a strongly negative Wigner function at the origin of phase space, $W(0,0)=-0.095$, Fig.~\ref{fig:wf} (e).  

\section{Conclusions}
In conclusion, we have demonstrated a balanced homodyne detector with high signal-to-noise ratio ($14.5$ dB), bandwidth ($80$ MHz) and unprecedented stability (measurement intervals of approximately 2 seconds between calibration) for use in  phase-sensitive detection, enabling shot-noise-limited quadrature measurements of ultrashort pulsed field modes. We have introduced the time-bandwidth product (TBP) for BHDs as a measure of the detector capacity to monitor high pulse repetition rate short-pulse sources,  reporting a TBP $\Delta f \Delta t = 1.6 \times 10^8$. The detector performance is further demonstrated by implementing QST of a coherent state and heralded single-photon state. We anticipate this detector will be useful for a variety of quantum and classical technologies such as quantum process tomography and phase-shift keying at low light levels. 

\begin{acknowledgments}
We are grateful for helpful discussions with M. E. Anderson, J. S. Lundeen, A. I. Lvovsky, D. McLellan, M. G. Raymer, and I. A. Walmsley. This work was supported by the University of Oxford John Fell Fund and EPSRC grant No. EP/E036066/1.
\end{acknowledgments}

\bibliography{bibliography}


\label{lastpage}

\end{document}